\documentclass[prd,twocolumn,showpacs,amsmath,amssymb,floatfix]{revtex4}
\usepackage{graphicx,color,dcolumn,booktabs}
\usepackage{longtable,lscape}
\usepackage{txfonts}
\usepackage{amssymb}
\usepackage{indentfirst}
\begin{document}
\title{$X(3915)$ and $X(4350)$ as new members in P-wave charmonium family}

\author{Xiang Liu$^{1,2}$\footnote{corresponding author}}\email{xiangliu@lzu.edu.cn}
\author{Zhi-Gang Luo$^3$}\author{Zhi-Feng Sun$^{1,2}$}

\affiliation{$^1$School of Physical Science and Technology, Lanzhou University, Lanzhou 730000,  China\\
$^2$Research Center for Hadron and CSR Physics,
Lanzhou University $\&$ Institute of Modern Physics of CAS, Lanzhou 730000, China\\
$^3$School of Physics, Peking University, Beijing 100871, China}

\date{\today}
\begin{abstract}
The analysis of the mass spectrum and the calculation of the strong decay of P-wave charmonium states strongly support to explain the newly observed $X(3915)$ and $X(4350)$ as new members in P-wave charmonium family, i.e., $\chi_{c0}^\prime$ for $X(3915)$ and $\chi_{c2}^{\prime\prime}$ for $X(4350)$. Under the P-wave charmonium
assignment to $X(3915)$ and $X(4350)$, the $J^{PC}$ quantum numbers of $X(3915)$ and $X(4350)$ must be
$0^{++}$ and $2^{++}$ respectively, which provide the important criterion to test P-wave charmonium explanation for $X(3915)$ and $X(4350)$ proposed by this letter. The decay behavior of the remaining two P-wave charmonium states with the second radial excitation is predicted, and experimental search for them is suggested.
\end{abstract}

\maketitle

Recently two new charmonium-like states $X(3915)$ and $X(4350)$ were released by the Belle Collaboration in the $\gamma\gamma$ fusion process \cite{:2009tx,:2009vs}.
$X(3915)$ is observed by the invariant mass spectrum of $J/\psi\omega$ in $\gamma\gamma\to J/\psi\omega$ channel.
The mass and width of $X(3915)$ are $M=3915\pm 3(\mathrm{stat})\pm2(\mathrm{sys})$ MeV and $\Gamma=17\pm10(\mathrm{stat})\pm 3(\mathrm{sys})$ MeV \cite{:2009tx}.
$X(4350)$, a new charmonium-like state found in the invariant mass spectrum of $J/\psi\phi$, is of $m=4350^{+4.6}_{-5.1}(\mathrm{stat})\pm0.7(\mathrm{sys})$ MeV and $\Gamma=13.3^{+17.9}_{-9.1}(\mathrm{stat})\pm4.1(\mathrm{sys})$ MeV \cite{:2009vs}.

Until now, Belle experiment has reported three charmonium-like states via the $\gamma\gamma$ fusion. Besides $X(3915)$ and $X(4150)$, $Z(3930)$ is a charmonium-like state observed in $\gamma\gamma\to D\bar{D}$, which is of mass $m=3929\pm(\mathrm{stat})5\pm2(\mathrm{syst})$ MeV and $\Gamma=29\pm10(\mathrm{stat})\pm2(\mathrm{syst})$ MeV \cite{Uehara:2005qd}. The angular distribution in the $\gamma\gamma$ center of mass frame shows $J^{PC}=2^{++}$, which indicates that $Z(3930)$ is a good candidate of $\chi_{c2}^\prime$, i.e., a charmonium with $n^{2s+1}J_L=2^3P_2$ \cite{Uehara:2005qd}.

\begin{center}
\begin{figure}[htb]
\scalebox{0.96}{\includegraphics{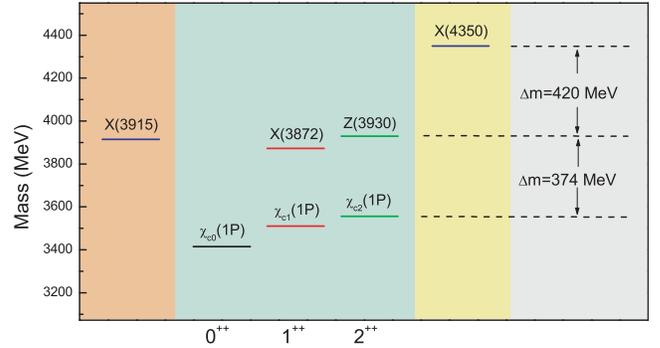}}
\caption{(Color online.) The established P-wave state charmonium states without the radial excitation \cite{Amsler:2008zzb} and the candidate
for the first radial excitation of P-wave state. A comparison between newly observed $X(4350)$ and the candidate of P-wave charmonium states is given. \label{review}}
\end{figure}
\end{center}

The observations of $X(3915)$ and $X(4350)$ ont only make the spectroscopy of charmonium-like state observed by the $\gamma\gamma$ fusion process become abundant, but also help us further reveal the underlying structure
of charmonium-like states observed by the $\gamma\gamma$ fusion.
Before illustrating the underlying structure of $X(3915)$ and $X(4350)$, one first gives a brief review of the established P-wave charmonium states or their possible candidates by Fig. \ref{review}, where $h_c(3525)$ (a P-wave state with spin 0) is not listed. Three P-wave states without the radiative excitation are $\chi_{c0}(3415)$, $\chi_{c1}(3510)$ and $\chi_{c2}(3556)$ \cite{Amsler:2008zzb}. For the first radial excitation of P-wave charmonium, the candidate for $0^{++}$ state $\chi_{c0}^\prime$ is still absent while $X(3872)$ \cite{Choi:2003ue,Kalashnikova:2005ui,Li:2009zu} and $Z(3930)$ \cite{Uehara:2005qd} can be recommended as $1^{++}$ state $\chi_{c1}^{\prime}$ and $2^{++}$ state $\chi_{c2}^\prime$, respectively.

Usually the $\gamma\gamma$ fusion process provides a good environment to create charmonium by $\gamma\gamma$ fusion into a pair of $c\bar{c}$. Although $X(3915)$ seems to be explained as an exotic state indicated in Ref. \cite{Nielsen:2009uh}, in this letter we adopt another point of view to explore whether the newly observed $X(3915)$ can fill in the blank of the remaining P-wave charmonium with the first radial excitation shown in Fig. \ref{review}. The mass of $X(3915)$ is consistent with the result of the potential model, which once predicted the mass of the first radial
excitation $\chi_{c0}^{\prime}$ is around 3916 MeV according to Godfrey-Isgur relativized potential model \cite{Barnes:2005pb}. Since there exists
the vanishing coupling of $X(3915)-D\bar D^*$ and the week interaction of $Z(3930)-D\bar{D}^*$ while $X(3872)$ with $J^{PC}=1^{++}$ interacting with $D\bar D^*$ via S-wave is very strong,
the coupled channel effect on bare $\chi_{c0}^{\prime}$ and $\chi_{c2}^{\prime}$ is weaker than that on bare $\chi_{c1}^{\prime}$ \cite{Kalashnikova:2005ui,Li:2009zu}, which explains why the mass difference between $X(3915)$ and $Z(3930)$
is smaller than that between $X(3915)$ and $X(3872)$.

By the comparison of $X(4350)$ with the existed P-wave states, one notices that the mass difference between $X(4350)$ and $Z(3930)$ is about 420 MeV, which is similar to that between $Z(3930)$ and $\chi_{c2}(1P)$. The regularity of the mass gaps existing $\chi_{c2}(1P)$, $Z(3930)$ and $X(4350)$ is consistent with the estimate from Resonance Spectrum Expansion (RSE) model, which indicates the mass gap between the states with the radial quantum numbers $n$ and $n+1$ is $380$ MeV \cite{vanBeveren:1979bd,vanBeveren:1982qb}. Thus, one further proposes that $X(4350)$ is as the second radial excitation of P-wave charmomium state. We also notice the prediction of the mass of $\chi_{c2}^{\prime\prime}$ in Ref. \cite{Barnes:2005pb} by Godfrey-Isgur relativized potential model, which is about $4337$ MeV. This value is consistent with the mass of $X(4350)$.

\renewcommand{\arraystretch}{1.3}
\begin{center}
\begin{table}[htb]
\begin{tabular}{c|c|c}\toprule[1pt]
State&Modes&Decay channels\\\midrule[1pt]
$\chi_{c0}^\prime$&$0^-+0^-$&$D\bar{D}$
\\\midrule[1pt]
&$0^-+0^-$&$D\bar D,\, D_s\bar{D}_s$\\
$\chi_{c0}^{\prime\prime}$&$1^-+1^-$&$D^*\bar{D}^*,\,D_s^*\bar{D}_s^*$\\
&$0^-+1^+$&$D\bar{D}_{1}(2430)+h.c.,\,D\bar{D}_{1}(2420)+h.c.$\\\midrule[1pt]
&$0^-+1^-$&$D\bar D^*+h.c.,\, D_s\bar{D}_s^*+h.c.$\\
$\chi_{c1}^{\prime\prime}$&$1^-+1^-$&$D^*\bar{D}^*,\,D_s^*\bar{D}_s^*$\\
&$0^-+0^+$&$D\bar{D}_0^*(2400)+h.c.,\,D_s\bar{D}_{s0}(2317)+h.c.$\\
&$0^-+1^+$&$D\bar{D}_{1}(2430)+h.c.,\,D\bar{D}_{1}(2420)+h.c.$\\\midrule[1pt]
&$0^-+0^-$&$D\bar D,\, D_s\bar{D}_s$\\
$\chi_{c2}^{\prime\prime}$&$0^-+1^-$&$D\bar D^*+h.c.,\, D_s\bar{D}_s^*+h.c.$\\
&$1^-+1^-$&$D^*\bar{D}^*,\,D_s^*\bar{D}_s^*$\\
&$0^-+1^+$&$D\bar{D}_{1}(2430)+h.c.,\,D\bar{D}_{1}(2420)+h.c.$\\
\bottomrule[1pt]
\end{tabular}
\caption{The allowed open-charm strong decays of $\chi_{c0}^\prime$ and $\chi_{cJ}^{\prime\prime}$ ($J=0,1,2$).
Here, we take 4350 MeV as the upper limit of the mass of $\chi_{cJ}^{\prime\prime}$.
$D_1(2420)$ is the $1^+$ state in the $T=(1^+, 2^+)$ doublet while $D_1(2430)$ is the $1^+$ state in the $S=(0^+, 1^+)$ doublet since as indicated in Ref. \cite{Luo:2009wu}. \label{decay} }
\end{table}
\end{center}

The decay modes of the created charmonium state from the $\gamma\gamma$ fusion include open-charm and hidden-charm decays, which are
the observed decay channel of $Z(3930)$ and $X(3915)/X(4350)$, respectively. For testing the proposal for the structure of $X(3915)$ and $X(4350)$, in the following, we further study open-charm decay of the radial excited P-wave charmonium $\chi_{c0}^{\prime}$ and $\chi_{cJ}^{\prime\prime}$  ($J=0,1,2$) by the Quark Pair Creation (QPC) model  \cite{Micu:1968mk,LeYaouanc:1977gm,LeYaouanc:1988fx}, which is a successful phenomenological model to calculate Okubo-Zweig-Iizuka (OZI) allowed strong decays of hadron.

The allowed decay modes of $\chi_{c0}^\prime$ and $\chi_{cJ}^{\prime\prime}$ are presented in Table \ref{decay}.
The QPC model provide us an effective approach to study the two-body strong decays of the radial excited P-wave charmonium $\chi_{c0}^{\prime}$ and $\chi_{cJ}^{\prime\prime}$.

\renewcommand{\arraystretch}{1.3}
\begin{center}
\begin{table}[htb]
\begin{tabular}{c|c|lc}\toprule[1pt]
State& Modes& \quad \quad\quad Partial wave amplitude\\\midrule[1pt]
$\chi_{c0}^\prime$&$0^-0^-$&$\mathcal{M}^{00}=\mathcal{F}\frac{\sqrt{2}}{3}\sqrt{E_AE_BE_C}\gamma
[2\mathcal{O}_{1,-1}-\mathcal{O}_{0,0}]$
\\\midrule[1pt]
&$0^-0^-$&$\mathcal{M}^{00}=\mathcal{F}\frac{\sqrt{2}}{3}\sqrt{E_AE_BE_C}\gamma
[2\mathcal{Q}_{1,-1}-\mathcal{Q}_{0,0}]$\\
&$1^-1^-$&$\mathcal{M}^{00}=-\mathcal{F}\sqrt{\frac{2}{27}}\sqrt{E_AE_BE_C}\gamma
[2\mathcal{Q}_{1,-1}-\mathcal{Q}_{0,0}]$\\
$\chi_{c0}^{\prime\prime}$&$0^-1^+(S)$&$\mathcal{M}^{11}=\mathcal{F}\sqrt{E_AE_BE_C}\gamma\{-\frac{\sqrt{2}\cos\theta}{3}
(2\mathcal{P}_{1,-1,0}$\\
&&$\quad\quad\quad-\mathcal{P}_{0,0,0})-\frac{2\sin\theta}{3}
(\mathcal{P}_{1,0,1}-\mathcal{P}_{0,1,1})\}$\\
&$0^-1^+(T)$&$\mathcal{M}^{11}=\mathcal{F}\sqrt{E_AE_BE_C}\gamma\{\frac{\sqrt{2}\sin\theta}{3}
(2\mathcal{P}_{1,-1,0}$\\
&&$\quad\quad\quad-\mathcal{P}_{0,0,0})-\frac{2\cos\theta}{3}
(\mathcal{P}_{1,0,1}-\mathcal{P}_{0,1,1})\}$\\\midrule[1pt]
&$0^-1^-$&$\mathcal{M}^{10}=-\mathcal{F}\frac{2}{3\sqrt{3}}\sqrt{E_AE_BE_C}\gamma
[2\mathcal{Q}_{1,-1}-\mathcal{Q}_{0,0}]$\\
&&$\mathcal{M}^{12}=\mathcal{F}\frac{2}{3\sqrt{6}}\sqrt{E_AE_BE_C}\gamma
[\mathcal{Q}_{1,-1}+\mathcal{Q}_{0,0}]$\\
&$1^-1^-$&$\mathcal{M}^{22}=\mathcal{F}\frac{2}{3}\sqrt{E_AE_BE_C}\gamma
[\mathcal{Q}_{1,-1}+\mathcal{Q}_{0,0}]$\\
$\chi_{c1}^{\prime\prime}$&$0^-0^+$&$\mathcal{M}^{01}=\mathcal{F}\frac{2}{3\sqrt{3}}\sqrt{E_AE_BE_C}\gamma
[\mathcal{P}_{1,-1,0}+\mathcal{P}_{1,0,1}]$\\
&$0^-1^+(S)$&$\mathcal{M}^{11}=\mathcal{F}\sqrt{E_AE_BE_C}\gamma\{\frac{\sqrt{2}\cos\theta}{3}
[\mathcal{P}_{0,1,1}-\mathcal{P}_{1,0,1}]$\\
&&$\quad\quad\quad+\frac{\sin\theta}{3}[\mathcal{P}_{0,0,0}+\mathcal{P}_{0,1,1}-\mathcal{P}_{1,-1,0}]\}$\\
&$0^-1^+(T)$&$\mathcal{M}^{11}=\mathcal{F}\sqrt{E_AE_BE_C}\gamma\{-\frac{\sqrt{2}\sin\theta}{3}
[\mathcal{P}_{0,1,1}-\mathcal{P}_{1,0,1}]$\\
&&$\quad\quad\quad+\frac{\cos\theta}{3}[\mathcal{P}_{0,0,0}+\mathcal{P}_{0,1,1}-\mathcal{P}_{1,-1,0}]\}$\\
\midrule[1pt]
&$0^-0^-$&$\mathcal{M}^{02}=\mathcal{F}\frac{2}{3\sqrt{5}}\sqrt{E_AE_BE_C}\gamma
[\mathcal{Q}_{1,-1}+\mathcal{Q}_{0,0}]$\\
&$0^-1^-$&$\mathcal{M}^{12}=\mathcal{F}\frac{2}{\sqrt{30}}\sqrt{E_AE_BE_C}\gamma
[\mathcal{Q}_{1,-1}+\mathcal{Q}_{0,0}]$\\
&$1^-1^-$&$\mathcal{M}^{20}=\mathcal{F}\frac{2}{3}\sqrt{\frac{2}{3}}\sqrt{E_AE_BE_C}\gamma
[2\mathcal{Q}_{1,-1}-\mathcal{Q}_{0,0}]$\\
$\chi_{c2}^{\prime\prime}$&$0^-1^+(S)$&$\mathcal{M}^{11}=\mathcal{F}\sqrt{E_AE_BE_C}\gamma\{\frac{\cos\theta}{15\sqrt{2}}
[4\mathcal{P}_{0,0,0}+6\mathcal{P}_{0,1,1}$\\
&&$+4\mathcal{P}_{1,-1,0}+6\mathcal{P}_{1,0,1}]+\frac{\sin\theta}{30}[6\mathcal{P}_{0,0,0}
+14\mathcal{P}_{0,1,1}$\\
&&$+6\mathcal{P}_{1,-1,0}+4\mathcal{P}_{1,0,1}]$\\
&$0^-1^+(T)$&$\mathcal{M}^{11}=\mathcal{F}\sqrt{E_AE_BE_C}\gamma\{-\frac{\sin\theta}{15\sqrt{2}}
[4\mathcal{P}_{0,0,0}+6\mathcal{P}_{0,1,1}$\\
&&$+4\mathcal{P}_{1,-1,0}+6\mathcal{P}_{1,0,1}]+\frac{\cos\theta}{30}[6\mathcal{P}_{0,0,0}
+14\mathcal{P}_{0,1,1}$\\
&&$+6\mathcal{P}_{1,-1,0}+4\mathcal{P}_{1,0,1}]$\\\bottomrule[1pt]
\end{tabular}
\caption{The partial wave amplitude of the open-charm decay of $\chi_{c0}^\prime$ and $\chi_{cJ}^{\prime\prime}$. Here, two $1^{+}$ charmed mesons in $S$ and $T$ doublets are the mixture of two basis $1^{1}P_1$ and $1^{3}P_1$, which indicates
$|1^{+}(S)\rangle=\cos\theta|1^1P_1\rangle+\sin\theta|1^3P_1\rangle$ and $|1^{+}(T)\rangle=-\sin\theta|1^1P_1\rangle+\cos\theta|1^3P_1\rangle$ \cite{Luo:2009wu}. One takes $\mathcal{F}=1/\sqrt 3$ obtained from the calculation of the flavor matrix element. In heavy quark limit,
one usually takes the mixing angle $\theta=-54.7^\circ$. According to the partial wave amplitude, the partial decay width is expressed
$\Gamma = \pi^2 \frac{{|\textbf{K}|}}{M_A^2}\sum_{JL}\big
|\mathcal{M}^{J L}\big|^2$
\cite{Micu:1968mk,LeYaouanc:1977gm,LeYaouanc:1988fx}.
\label{partial}}
\end{table}
\end{center}

The transition matrix element of the process $A(c(1)\bar{c}(2))\to B(c(1)\bar{q}(3))+C(\bar{c}(2)q(4))$ in the center of mass frame of charmonium $A$ is written as
$\langle BC|\mathcal{T}|A \rangle=\delta^3(\mathbf{K}_B+\mathbf{K}_C)M^{M_{J_A}M_{J_B}M_{J_C}}(\mathbf{K})$,
where the transition operator $\mathcal{T}$ in the QPC model reads as
\begin{eqnarray}
\mathcal{T}&=& - 3 \gamma \sum_m\: \langle 1\;m;1\;-m|0\;0 \rangle\,
\int\!{\rm d}{\textbf{k}}_3\; {\rm
d}{\textbf{k}}_4 \delta^3({\textbf{k}}_3+{\textbf{k}}_4) \nonumber\\&&\times{\cal
Y}_{1m}\left(\frac{{\textbf{k}}_3-{\textbf{k}_4}}{2}\right)\;
\chi^{3 4}_{1, -\!m}\; \varphi^{3 4}_0\;\,
\omega^{3 4}_0\; d^\dagger_{3i}({\textbf{k}}_3)\;
b^\dagger_{4j}({\textbf{k}}_4)\,,\nonumber
\end{eqnarray}
where $i$ and $j$ denote the SU(3) color indices of the created quark
and anti-quark from the vacuum. $\varphi^{34}_{0}=(u\bar u +d\bar d +s \bar
s)/\sqrt 3$ and $\omega_{0}^{34}=\delta_{\alpha_3\alpha_4}/\sqrt 3\,(\alpha=1,2,3)$ means flavor and
color singlets, respectively. $\chi_{{1,-m}}^{34}$ is a triplet
state of spin. $\mathcal{Y}_{\ell m}(\mathbf{k})\equiv
|\mathbf{k}|^{\ell}Y_{\ell m}(\theta_{k},\phi_{k})$ is the
$\ell$th solid harmonic polynomial. Dimensionless
constant $\gamma$ is the strength of the quark pair creation from the
vacuum and can be extracted by fitting the data. In this letter, $\gamma=6.3$ \cite{Godfrey:1986wj}. The strength of $s\bar{s}$ creation satisfies
$\gamma_{s}=\gamma/\sqrt{3}$ \cite{LeYaouanc:1977gm}.
The helicity amplitude $M^{M_{J_A}M_{J_B}M_{J_C}}(\mathbf{K})$ is extracted by the transition matrix element, which is related to the partial wave amplitude by \cite{Jacob:1959at}
\begin{eqnarray}
{\mathcal{M}}^{J L}(A\rightarrow BC) &=&
\frac{\sqrt{2 L+1}}{2 J_A +1} \!\! \sum_{M_{J_B},M_{J_C}} \langle
L 0 J M_{J_A}|J_A  M_{J_A}\rangle \nonumber\\&&\times\langle
J_B M_{J_B} J_C  M_{J_C} | J M_{J_A} \rangle \mathcal{M}^{M_{J_A}
M_{J_B} M_{J_C}}({\textbf{K}}),\nonumber\label{JB}
\end{eqnarray}
where $\mathbf{J}=\mathbf{J}_B+\mathbf{J}_C$ and
$\mathbf{J}_{A}+\mathbf{J}_{P}=\mathbf{J}_{B}+\mathbf{J}_C+\mathbf{L}$. A detailed review of the QPC model was illustrated in Ref. \cite{Luo:2009wu}.
The partial wave amplitude corresponding to the open-charm decays shown in Table \ref{decay} is presented in Table \ref{partial}.
The concrete expressions of $\mathcal{O}_{i,j}$, $\mathcal{Q}_{i,j}$ and $\mathcal{P}_{i,j,k}$ are extracted from the spatial integral $I^{M_{L_A},m}_{M_{L_B},M_{L_C}}(\textbf{K})$, which describes the overlap of the
initial meson ($A$) and the created pair with the two final mesons
($B$ and $C$). Here, the harmonic oscillator (HO) wave function $\Psi_{n_r\ell m}(\mathbf{k}) = \mathcal{R}_{n_r\ell}(R,\mathbf{k})\mathcal{Y}_{n_r\ell m}(\mathbf{k})$ is involved in the calculation of the spatial integral. Parameter $R$ in the HO wave function is obtained by reproducing the realistic root mean square (RMS)
radius by solving the schr\"{o}dinger equation with the linear
potential \cite{Godfrey:1986wj}. The relevant mass values are taken in PDG \cite{Amsler:2008zzb}. We take $R=1.52,\,1.41,\,1.85,\,1.69,\,1.85,\,1.75,\,2.00,\,2.00$ GeV$^{-1}$ corresponding to
$D$, $D_s$, $D^*$, $D_s^*$, $D_{0}^*(2400)$, $D_{s0}^*(2317)$, $D_{1}(2430)$ and $D_1(2420)$, respectively.
Besides, other parameter inputs include $m_c=1.6$ GeV, $m_u=m_d=0.22$ GeV and $m_s=0.419$ GeV.

\begin{center}
\begin{figure}[htb]
\scalebox{0.85}{\includegraphics{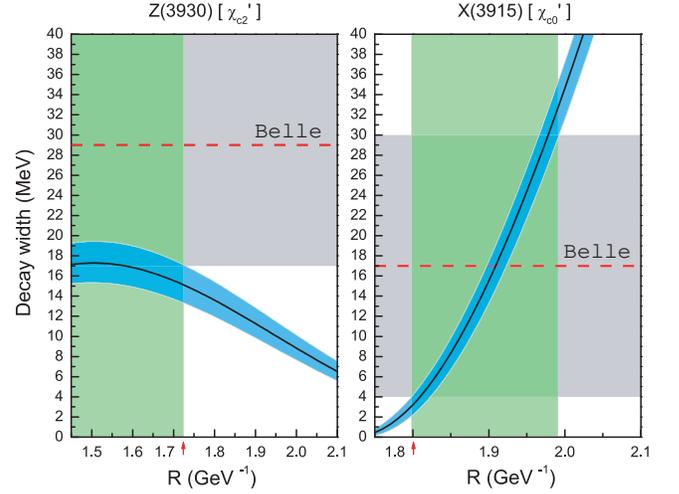}}
\caption{(Color online.) The dependence of the decay width of $Z(3930)$ and $X(3915)$ on $R$ under $\chi_{c2}^\prime$ and $\chi_{c0}^\prime$ assignment for $Z(3930)$ and $X(3915)$, respectively. Here, red dash line with grey band denote the central value for the error of total width
of $X(3915)$ and $Z(3930)$ measured by Belle \cite{:2009tx,Uehara:2005qd}. The green band denotes the region of $R$ resulting in the theoretical values consistent with Belle data. The solid lines with blue error bands are our calculation result. \label{chic0}}
\end{figure}
\end{center}

In the right diagram of Figs. \ref{chic0}, one presents the total width of $\chi_{c0}^\prime$ with the variation of $R$.
The node effect from the wave function of higher radial excited states results in
the decay width calculated by the QPC model being dependent on the $R$ value. When taking $R=1.80\sim 1.99$ GeV$^{-1}$, the obtained total open-charm decay width of $\chi_{c0}^\prime$ falls in the range of total width of $X(3915)$ released by Belle \cite{:2009tx} (the calculation result with the typical value $R=1.92$ GeV$^{-1}$ corresponds to the central value of the width of $X(3915)$).
As a $\chi_{c2}^\prime$ charmonium state \cite{Uehara:2005qd}, $Z(3930)$ can be as a realistic test of the reasonability of the range of $R$ for $\chi_{c0}^\prime$. Our theoretical result
of the open-charm decay of $\chi_{c2}^\prime$ dependent on $R$ and the comparison of the calculation result with Belle data of $Z(3930)$ \cite{Uehara:2005qd}, which are shown in the left diagram of Figs. \ref{chic0},
indicate that the upper limit of $R$ for $\chi_{c2}^\prime$ is very close to the lower limit of $R$ for $\chi_{c0}^\prime$ as marked by the red arrows in Fig. \ref{chic0}, which further
shows the reliability of investigating $\chi_{c1}^\prime$ open-charm decay in the range of $R=1.80\sim 1.99$ GeV$^{-1}$. Thus, explaining $X(3915)$ as a $\chi_{c0}^\prime$ charmonium is tested through the open-charm decay of $\chi_{c0}^\prime$.

Considering an exotic $D^*\bar{D}^*$ molecule explanation to $X(3915)$ suggested in Ref. \cite{Nielsen:2009uh}, one proposes the experimental study of the open-charm decay $D\bar{D}$ to be as a best way to distinguish between the exotic and the conventional states for the controversial $X(3915)$ since the $D\bar{D}$ decay of $X(3915)$ under the $D^*\bar{D}^*$ molecule assignment occurs via hadronic loop effect as indicated in Ref. \cite{Liu:2009ei}, which
results in the decay width of $X(3915)\to D\bar{D}$ under the $D^*\bar{D}^*$ molecule assignment being far smaller than that under $\chi_{c0}^\prime$ explanation to $X(3915)$.

\begin{center}
\begin{figure}[htb]
\scalebox{0.87}{\includegraphics{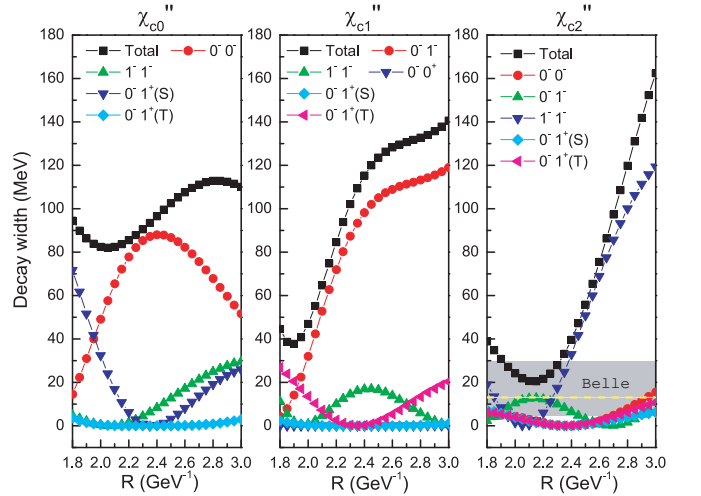}}
\caption{(Color online.) The variation of the decay width of $\chi_{cJ}^{\prime\prime}$ ($J=0,1,2$) with $R$ value. Here, we set the upper limit of the masses of P-wave states
with the second radial excitation as 4.35 GeV. The yellow dash line and shaded grey band shown in diagram of $\chi_{c2}^{\prime\prime}$ denote the central value for the error of total width
of $X(3915)$ measured by Belle \cite{:2009vs}. \label{p-wave}}
\end{figure}
\end{center}

The results of the open-charm decays of $\chi_{cJ}^{\prime\prime}$ ($J=0,1,2$) are presented in Fig. \ref{p-wave} via scanning the parameter space $R=1.8\sim 3.0$ GeV$^{-1}$, which is due to the radius $R$ of $\chi_{cJ}^{\prime\prime}$ being fatter than that of $\chi_{cJ}^\prime$. We need to emphasize that the $\gamma\gamma$ fusion process determines the most possible quantum number of $X(4350)$ to be $0^{++}$ or $2^{++}$, which makes us choose $X(4350)$ as the candidate of $\chi_{c0}^{\prime\prime}$ or $\chi_{c2}^{\prime\prime}$ and fully exclude $\chi_{c1}^{\prime\prime}$ assignment to $X(4350)$ \cite{:2009vs}. The open-charm decays of $\chi_{c0}^{\prime\prime}$ and $\chi_{c2}^{\prime\prime}$, two candidates of $Y(4350)$, display different behaviors as illustrated in the left and right diagrams of Fig. \ref{p-wave}.  The total open-charm decay of $\chi_{c2}^{\prime\prime}$ with $R=1.9\sim 2.3$ GeV$^{-1}$ is well consistent with Belle data \cite{:2009vs} as shown in the right diagram of Fig. \ref{p-wave} while the total open-charm decay of $\chi_{c0}^{\prime\prime}$ is far away from the Belle data \cite{:2009vs}, which shows that we can fully exclude $\chi_{c0}^{\prime\prime}$ explanation for $X(4350)$ and finally establish $X(4350)$ as a good candidate of $\chi_{c2}^{\prime\prime}$. Our numerical result demonstrates that $1^-1^-/0^-1^-$ ($D\bar{D}^*+h.c., D_s\bar{D}_s^*+h.c.,\, D^*\bar{D}^*,\,D_s^*\bar{D}_{s}^*$) are the dominant decay channels for $X(4350)$.

Meanwhile, the predicted properties of the remaining two P-wave charmonium states with the second radial excitation can be as the guidance of experimental search for $\chi_{c0}^{\prime\prime}$ and $\chi_{c1}^{\prime\prime}$. The predicted total decay width of $\chi_{c0}^{\prime\prime}$ is around $82\sim 110$ MeV corresponding to $R=2.0\sim 3.0$ GeV$^{-1}$, which is not strongly dependent on the $R$ values. As the dominant decay mode of $\chi_{c0}^{\prime\prime}$,
$0^-0^-$ channel including $D\bar{D}$ and $D_s^+D_s^-$ is a golden channel to find $\chi_{c0}^{\prime\prime}$. For $\chi_{c1}^{\prime\prime}$, its total decay width is of large span from $47$ MeV to $140$ MeV corresponding to $R=2.0\sim 3.0$ GeV$^{-1}$. Among its partial decay channels, $0^-1^-$ channel is always the main decay channel of $\chi_{c1}^{\prime\prime}$ under taking different values of $R$, which indicates that $D\bar{D}^*+h.c.$ and $D_s^+D_s^{*-}+h.c.$ can be as the suggested decay channel of searching for $\chi_{c1}^{\prime\prime}$. The detail of the open-charm decay behaviors of $\chi_{cJ}^{\prime\prime}$ are listed in Fig. \ref{p-wave}.

In summary, the newly observed $X(3915)$ and $X(4350)$ are firstly explained as $\chi_{c0}^{\prime}$ and $\chi_{c2}^{\prime\prime}$ extremely well respectively by analyzing
the mass spectrum of P-wave charmonium family as well as by calculating the open-charm strong decay of $X(3915)$ and $X(4350)$, which
are consistent with the existed experimental findings. Just because of our explanations to $X(3915)$ and $X(4350)$, the spectroscopy of P-wave charmonium becomes abundant.
Under the assignment of P-wave charmonium to $X(3915)$ and $X(4350)$, the $J^{PC}$ quantum numbers of $X(3915)$ and $X(4350)$ will be definite, i.e. $J^{PC}=0^{++}$ for $X(3915)$ and $J^{PC}=2^{++}$ for $X(4350)$, which provides the powerful criterion to test the P-wave charmonium assignment for $X(3915)$ and $X(4350)$ since the experimental analysis of the angular distribution of $X(3915)$ and $X(4350)$
can give the concrete information of their quantum numbers with model independent.
Additionally, in this work, we also predict the decay behaviors of the two remaining second radial excited P-wave charmonium states $\chi_{c0}^{\prime\prime}$ and $\chi_{c1}^{\prime\prime}$. These findings are expected to be revealed in future experiment.

This study can be extended to include the theoretical study of the hidden-charm decay, the radiative decay and
the double-photon decay of $X(3915)$ and $X(4350)$, which will provide us valuable information of their
underlying structure.

Note added: Recently, a theoretical work using QCD sum rule \cite{Albuquerque:2010fm} shows that it is not possible to describe the $X(4350)$ structure as a $1^{-+} D_s^*{D}_{s0}^*$ molecular state, which
supports our effort to explain $X(4350)$ under the conventional charmonium to some extent.

{\bf Acknowledgement}
This project is supported by the National Natural Science Foundation of
China (NSFC) under Contracts No. 10705001; the Foundation for
the Author of National Excellent Doctoral Dissertation of P.R. China
(FANEDD) under Contracts No. 200924; the Doctoral Program Foundation of Institutions of
Higher Education of P.R. China under Grant No. 20090211120029; the Program for New Century Excellent Talents in University (NCET) by Ministry of Education of P.R. China.

\end{document}